# Waste-to-Energy-Coupled AI Data Centers: Cooling Efficiency and Grid Resilience

Qi He
Google
USA

Chunyu Qu
Dun & Bradstreet Inc.
USA

***Abstract*—AI data center expansion is increasingly constrained by grid interconnection capacity, driven in part by electricity-intensive cooling demand. This paper evaluates a coupled Waste-to-Energy (WtE) and AI data center configuration in which low-grade WtE thermal output drives absorption cooling, displacing grid-supplied cooling electricity. The system is modeled as a transparent input–output boundary and benchmarked against a conventional grid-powered cooling baseline. Results show that net grid-capacity relief is governed by three first-order factors: cooling coverage of IT heat load, parasitic auxiliary power, and distance-dependent thermal delivery losses. These factors define a break-even siting corridor beyond which capacity benefits vanish. The analysis shows that WtE-driven cooling can reduce grid imports under realistic conditions, enabling tighter integration of energy-generation resources with data centers.**

## I. Introduction

AI data-center (AIDC) expansion is increasingly constrained not by model architectures but by the joint availability of deliverable energy services: reliable electricity and scalable heat rejection. Recent assessments note that DCs already constitute a material electricity load and that AI-driven growth could sharply increase demand over the next decade [1], [2]. In practice, the binding constraints are local and physical, including interconnection limits, long equipment lead times, and community caps on thermal and water footprints. What becomes scarce is not compute in the abstract, but the capacity to deploy compute within a feasible electricity–cooling envelope.

Cooling sits at the center of that envelope because it governs both operating cost and feasible power density. Conventional mechanical cooling uses high-grade electricity to move low-grade heat to ambient. Even with containment, economizers, and advanced controls, it remains anchored to compressor and fan work that scales with thermal load. The industry's shift toward liquid-assisted architectures improves near-chip heat transfer, yet many deployments still depend, partially or fully, on electricity-driven chillers, cooling towers, dry coolers, or hybrid systems for ultimate heat rejection [3]. Cooling therefore remains a first-order driver of marginal electricity demand and siting feasibility.

This paper proposes a complementary pathway: treat cooling as an energy service that can be supplied through grade matching, using low-grade thermal energy to deliver cooling via thermally driven cycles and displace compressor electricity. Waste-to-energy (WtE) from municipal solid waste is a salient opportunity because modern regulated plants produce a steady low-grade thermal stream suitable for absorption cooling, while also mitigating landfill methane emissions through waste diversion [4], [5]. WtE deployment is often perceived as "dirty," but contemporary facilities operate under stringent pollutant-control regimes embedded in enforceable standards across major jurisdictions [6]–[9]. Large-sample evidence further treats emissions control and resource recovery as measurable performance levers rather than fixed constraints [10]. Conditional on a modern compliant WtE plant, the decision-relevant question becomes: can coupling WtE thermal output to AIDC cooling deliver economically meaningful electricity displacement under realistic spatial and parasitic constraints?

This study addresses the question with a transparent thermoeconomic screening framework that models the coupled WtE–AIDC as an input–output boundary, benchmarked against a conventional grid-powered cooling baseline. The analysis shows that net grid-capacity relief is governed by three first-order determinants: cooling coverage of IT heat load, parasitic auxiliary electricity, and distance-driven thermal delivery decay. These determinants yield siting-ready break-even corridor conditions for urban deployment. By separating first-order feasibility from second-order component optimization, the paper provides decision-ready criteria for integrating an energy-generation resource with an emerging high-impact load. It also outlines an accounting interface to operational efficiency metrics and a computable Levelized Cost of Computing (LCOC) without re-deriving full lifecycle inventories.

Recent post-2022 work underscores why this screening perspective matters. System-level life-cycle evidence indicates that advanced cooling choices can materially shift the energy and water footprint of cloud infrastructure, making cooling technology a consequential lever rather than a back-end engineering detail [11]. In parallel, technoeconomic studies show that thermal-side interventions, such as reservoir-scale thermal energy storage for cooling, can achieve large electricity displacement at low levelized cost, and that ultra-low-grade-heat-driven absorption configurations can operate at driving temperatures near 50 °C for DC cooling applications [12], [13].

A second gap concerns how candidate thermal sources are characterized under compliance constraints. Recent cross-plant evidence using independently validated environmental statements shows that modern WtE facilities can achieve pollutant levels within best-available-technique associated emission levels and that performance varies systematically with identifiable technical features [14]. Bridging these two strands, cooling innovation and regulated thermal supply, motivates a siting-first framework that is explicit about boundaries, parasitics, and distance decay. The present study develops that bridge for WtE-coupled AIDC cooling and derives corridor-style feasibility conditions that can be evaluated with auditable inputs.

More broadly, recent work on LLM serving efficiency, retrieval-augmented generation, prompt behavior, domain-specific AI systems, and graph-based prediction highlights the growing importance of system-aware AI deployment across infrastructure and applied decision contexts [15]–[22].

## II. INTEGRATED SYSTEM TOPOLOGY AND GRADE

his section presents the integrated WtE–AIDC concept using an input–output accounting lens. We abstract from equipment-level design and focus on the minimum set of energy-service flows that determine feasibility and value: waste-derived primary energy, electricity, recoverable heat, delivered cooling, grid purchases, and parasitic requirements. Figure 1 defines the topology and accounting boundary; Figure 2 summarizes the corresponding cascade allocation of waste-derived services.

### A. *System topology, boundary conditions, and the counterfactual benchmark*

Figure 1 models three nodes linked by energy-service pathways. The WtE plant converts municipal solid waste $W$ with heating value $LHV$ into electricity $E^{wte}$ and recoverable heat $Q^{wte}$. Electricity can supply the AI data center (AIDC) directly (or equivalently offset grid purchases). Recoverable heat is routed through a coupling link that converts delivered heat into cooling via absorption chilling. The coupling is characterized by an absorption performance parameter $COP_a$, a distance-dependent thermal delivery factor $\eta_{tr}(L)$, and parasitic electricity demand $W_{par}(L)$ capturing pumping and auxiliary loads.

The AIDC consumes IT power $P_{IT} = \rho P_{IT}^{max}$ and requires cooling $Q_{req} = \gamma P_{IT}$ to deliver compute service $K$. Grid electricity $E^{grid}$ at price $p^e$ supplies IT load, fixed overhead, and any residual cooling not covered by the heat-driven pathway. The dashed connection in Figure 1 denotes the standalone counterfactual: absent coupling, cooling is met entirely by mechanical chilling with coefficient $COP_m$, implying cooling electricity $Q_{req}/COP_m$. In the integrated system, heat-driven cooling partially or fully displaces this component, reducing purchased electricity and improving operational metrics.

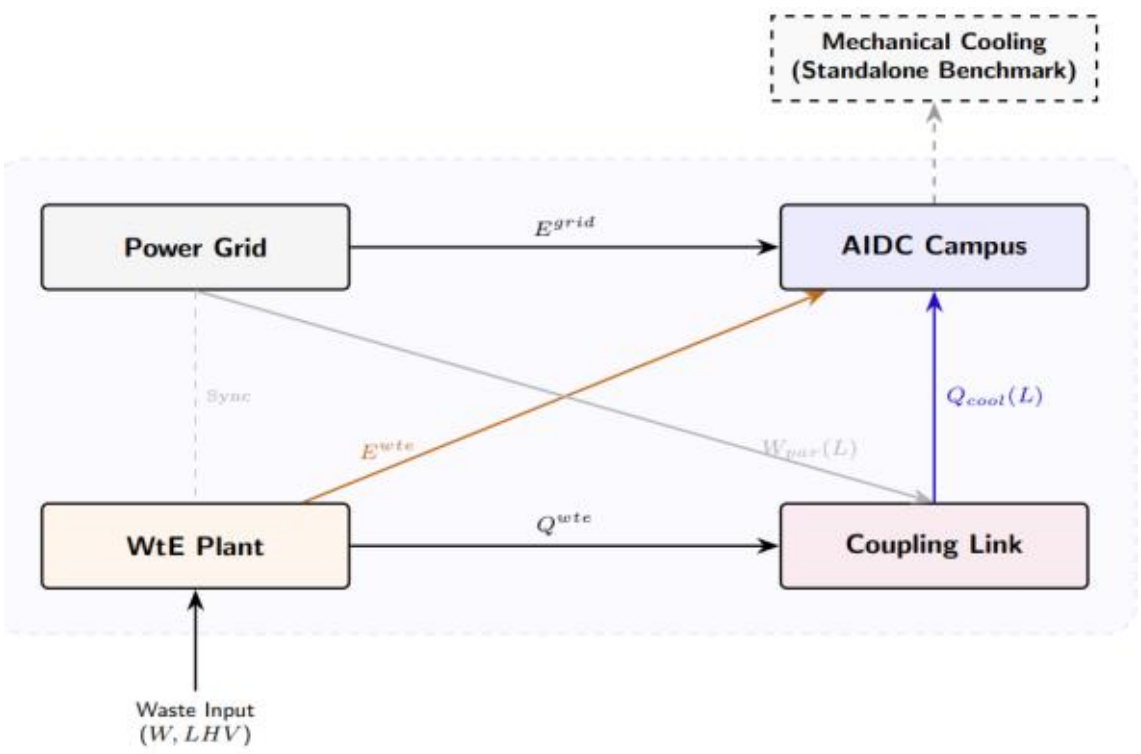

Figure 1. System architecture and energy-service flow topology of the integrated WtE and AI DC synergetic framework.

### B. *Grade matching and cascade utilization*

Figure 2 summarizes the core mechanism: grade matching through cascade utilization. Conventional WtE evaluations often emphasize electricity output while treating residual heat as low value. In contrast, AIDC operation requires continuous cooling. The integrated design redirects low-grade thermal output $Q^{wte}$ to a high-demand sink (cooling), producing an electricity-offsetting service. Conceptually, the same waste-energy input yields a portfolio of useful services, electricity and cooling, where cooling is valuable because it avoids mechanical cooling electricity that would otherwise be purchased from the grid.

## III. THERMOECONOMIC MODEL WITH MINIMAL THERMODYNAMIC CONSTRAINTS

This section formalizes a screening-level thermoeconomic model for the coupled WtE–AIDC architecture. The objective is not equipment design, but a conservation-consistent accounting boundary with the minimum physical constraints needed for credible comparison against a standalone grid-powered cooling baseline and for the comparative statics in Section 4.

Figure 2. Cascade energy accounting and service fulfillment topology of the integrated WtE-AIDC nexus

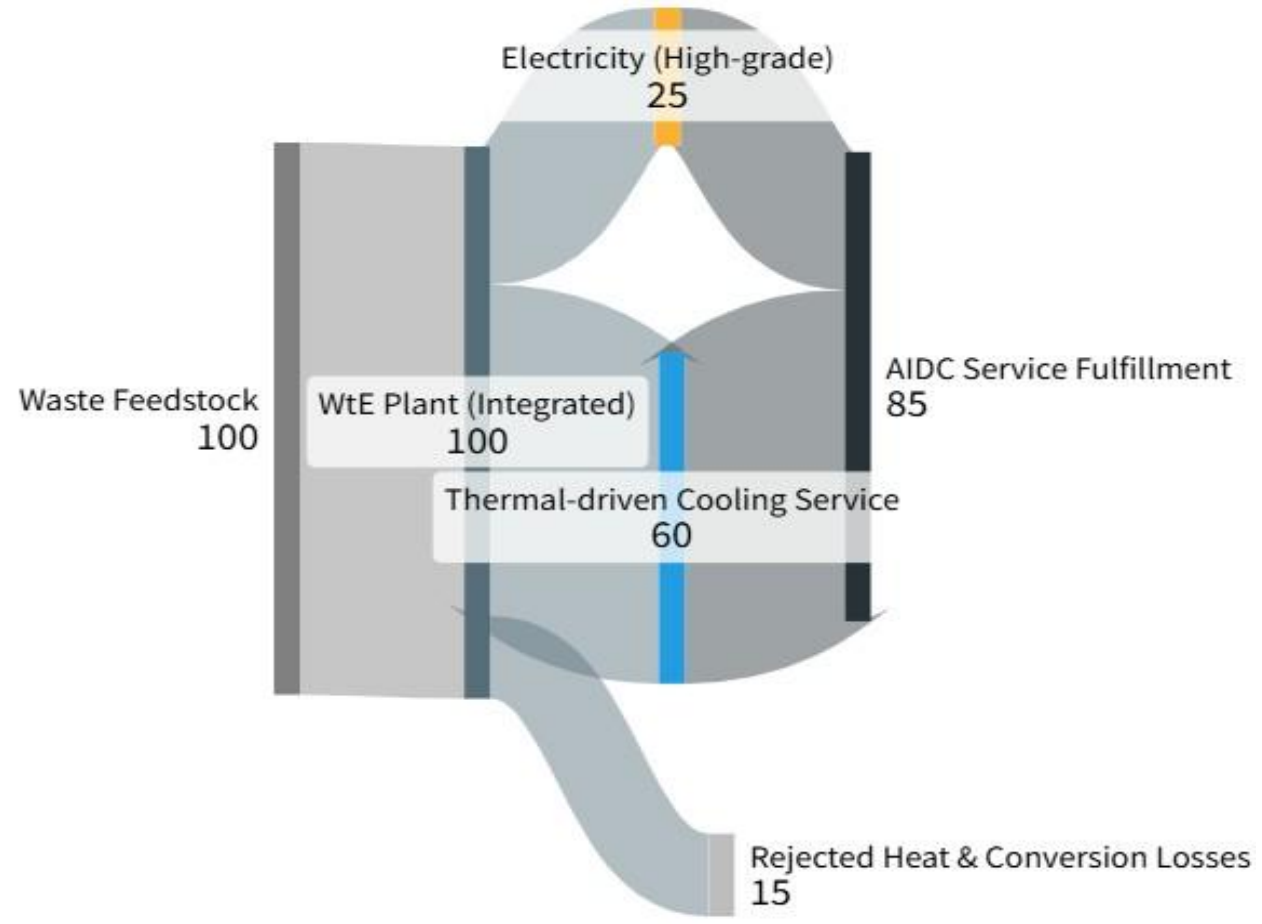

## A. Notation and Boundary Assumptions

All quantities are steady-state rates (dot notation). The coupled system boundary includes: (i) a WtE plant converting waste chemical energy to electricity and recoverable heat, (ii) a thermal bridge delivering usable heat to the DC, and (iii) absorption cooling that converts delivered heat to cooling service. We distinguish (a) facility electricity accounting at the DC and (b) system accounting over the coupled boundary to prevent misleading "efficiency gains" that arise solely from shifting cooling energy from electricity to heat.

LIST OF SYMBOLS AND THERMODYNAMIC PARAMETERS

| Symbol | Definition | Units |
|---|---|---|
| $\dot{m}_w$ | Waste mass flow rate | kg/s |
| $LHV$ | Lower heating value of waste | MJ/kg |
| $\dot{E}_{in}$ | Primary energy input from waste ($\dot{m}_w \cdot LHV$) | MW |
| $\eta_c$ | Effective combustion/boiler utilization efficiency | – |
| $\eta_e$ | Net electrical conversion efficiency (WtE) | – |
| $\alpha_h$ | Recoverable heat fraction of non-electric output (heat recovery factor) | – |
| $\dot{W}_e$ | Net electric output from WtE | MW |
| $\dot{Q}_h$ | Recoverable thermal output available for downstream use | MW |
| $L$ | Separation distance (WtE to DC) / thermal bridge length | km (or m) |
| $\eta_{tr}(L)$ | Effective thermal delivery factor (reduced-form) | – |
| $\beta$ | Thermal decay coefficient in $\eta_{tr}(L)$ | 1/km |
| $COP_{abs}$ | Absorption chiller coefficient of performance | – |
| $COP_m$ | Mechanical (electric) chiller COP | – |
| $\dot{Q}_{cool}$ | Cooling rate supplied to DC | MW |
| $\dot{W}_{IT}$ | Data center IT electrical load | MW |
| $\gamma$ | Cooling requirement per IT load ($\dot{Q}_{req} = \gamma \dot{W}_{IT}$) | – |
| $\dot{W}_{aux}$ | Non-cooling auxiliary DC load (fans, UPS losses, lighting) | MW |
| $\dot{W}_{par}$ | Parasitic work for thermal delivery & absorption (pumps, controls) | MW |
| $T_0$ | Ambient reference temperature | K |
| $T_c$ | Chilled-water / refrigeration temperature | K |
| $\phi_c$ | Cooling exergy factor $= T_0/T_c - 1$ | – |

## B. Minimal Energy-Flow Equations

*(1) Waste-to-energy conversion*

The primary energy input rate from waste is:

$$\dot{E}_{in} = \dot{m}_w \cdot LHV \tag{1}$$

Net electricity output is modeled as:

$$\dot{W}_e = \eta_e \, \eta_c \, \dot{E}_{in} \tag{2}$$

Recoverable thermal output available for downstream use is modeled as a fraction of the non-electric useful energy:

$$\dot{Q}_h = \alpha_h \left( \eta_c \dot{E}_{in} - \dot{W}_e \right) \tag{3}$$

*(2) Thermal delivery (spatial feasibility constraint)*

Thermal delivery to the DC is constrained by distance $L$ through a reduced-form delivery factor:

$$\dot{Q}_{del} = \eta_{tr}(L) \, \dot{Q}_h, \eta_{tr}(L) = \exp\left(-\beta L\right) \tag{4}$$

This reduced-form factor compactly represents aggregate losses and operational constraints without committing to a specific piping design, which is appropriate for siting-ready screening.

*(3) Absorption cooling transformation*

Delivered heat converted to cooling via absorption chilling:

$$\dot{Q}_{cool} = COP_{abs} \, \dot{Q}_{del} \tag{5}$$

The DC's cooling requirement is parameterized as:

$$\dot{Q}_{req} = \gamma \dot{W}_{IT} \tag{6}$$

Define the heat-driven coverage share:

$$f \equiv \min\{1, \frac{\dot{Q}_{cool}}{\dot{Q}_{req}}\} \tag{7}$$

Residual cooling demand $(1-f)\dot{Q}_{req}$ is met by mechanical chilling, requiring electricity:

$$\dot{W}_{cool,m} = \frac{(1-f)\dot{Q}_{req}}{COP_m} \tag{8}$$

Finally, parasitic work $\dot{W}_{par}$(pumps, controls, circulation) is modeled as a transparent reduced-form load:

$$\dot{W}_{par} = \kappa_0 + \kappa_1 \dot{W}_{IT} + \kappa_2 L \tag{9}$$

where $\kappa_0, \kappa_1, \kappa_2$can be scenario-calibrated, keeping the model auditable while preserving the focus on system accounting.

## C. System-Consistent Performance Metrics

*(1) Extended Power Usage Effectiveness (PUE) - auditable under heat-driven cooling*

Conventional PUE is electricity-only and becomes non-comparable once cooling is supplied thermally. We therefore define two complementary metrics:

(i) Electric PUE (facility electricity only):

$$PUE_{elec} \equiv \frac{\dot{W}_{IT} + \dot{W}_{aux} + \dot{W}_{cool,m} + \dot{W}_{par}}{\dot{W}_{IT}} \tag{10}$$

This metric is directly relevant to grid capacity and electric operating expenses (OpEx).

*(2) Boundary-consistent "system PUE"* (energy-service intensity). Because cooling may be supplied by delivered heat, we define a boundary-consistent ratio that counts heat delivered into the coupled boundary:

$$PUE_{sys} \equiv \frac{\dot{W}_{IT} + \dot{W}_{aux} + \dot{W}_{cool,m} + \dot{W}_{par} + \dot{Q}_{del}}{\dot{W}_{IT}} \tag{11}$$

$PUE_{sys}$is not ISO/IEC PUE and is not intended for cross-study benchmarking; it is introduced solely for conservation-

consistent bookkeeping when cooling is supplied via non-electric energy services.

*(3) Systemic Exergy Efficiency ($\eta_{ex}$).* To distinguish energy quality, we treat waste chemical energy as the primary exergy input (screening approximation):

$$Ex_{in} \equiv \eta_c \dot{E}_{in} \tag{12}$$

Useful exergy output includes electricity and the exergy of delivered cooling. The exergy of a cooling effect at temperature $T_c$relative to ambient $T_0$is:

$$\phi_c \equiv \frac{T_0}{T_c} - 1, Ex_{cool} \equiv \phi_c \dot{Q}_{cool} \tag{13}$$

Thus total useful exergy output is:

$$Ex_{out} \equiv \dot{W}_e + Ex_{cool} \tag{14}$$

Define systemic exergy efficiency:

$$\eta_{ex} \equiv \frac{Ex_{out}}{Ex_{in}} = \frac{\dot{W}_e + \phi_c \dot{Q}_{cool}}{\eta_c \dot{E}_{in}} \tag{15}$$

$\eta_{ex}$is a quality-adjusted utilization index, which increases when low-grade heat is converted into a service (cooling) that offsets high-value electricity use, i.e., when the system reduces exergy destruction through better grade matching.

### D. A sufficient condition for thermoeconomic superiority

Proposition: Holding $\dot{W}_{IT}$fixed, a sufficient condition for the coupled system to weakly dominate the standalone architecture in systemic exergy utilization is $\phi_c \dot{Q}_{cool} \geq \dot{W}_{par}$. Substituting (4)–(5), this becomes:

$$\phi_c \, COP_{abs} \, \eta_{tr}(L) \, \dot{Q}_h \geq \dot{W}_{par} \tag{16}$$

Proof sketch (intuition): In the standalone case, recoverable heat $\dot{Q}_h$ is predominantly rejected, contributing no useful exergy output. Under coupling, the same heat is converted into cooling whose exergy content is $\phi_c \dot{Q}_{cool}$. The coupled system improves exergy utilization whenever this quality-adjusted cooling contribution exceeds the additional parasitic work required to deliver and convert the heat.

## IV. Simulation Design and Comparative Statics

This section quantifies the operational and thermoeconomic implications of integrating a WtE plant with an AIDC through a heat-driven cooling pathway. The simulations are intentionally "minimum-physics, maximum-accountability": we impose conservation-consistent thermodynamic boundaries and parameterize only the margins that govern feasibility and value, IT load utilization, feedstock quality, and spatial heat-delivery constraints. Figures 1 and 2 summarize the baseline performance and quality-adjusted efficiency gains; Figures 4.3 and 4.5 provide comparative statics on (i) waste heating value and (ii) distance-driven thermal decay and parasitic requirements.

### A. Baseline configuration and calibration

We consider a representative AIDC with maximum IT capacity $P_{IT}^{max}$and utilization $\rho \in (0,1]$, so that $P_{IT} = \rho P_{IT}^{max}$. The cooling requirement is modeled as proportional to IT load, $Q_{req} = \gamma P_{IT}$, where $\gamma$ captures site and design conditions. In the standalone benchmark, cooling is supplied electrically via mechanical chillers with coefficient of performance $COP_m$. In the integrated configuration, recoverable WtE thermal output drives an absorption chiller with $COP_a$, delivered through a heat-transport link with exponential decay $\eta_{tr}(L) = \exp(-\beta L)$, where $L$is the separation distance and $\beta$is an effective thermal attenuation rate.

Let $Q_{cool}(L)$ be the delivered cooling potential from WtE heat at distance $L$. Define the thermal coverage ratio $f(L) \equiv \min\left\{1, \frac{Q_{cool}(L)}{Q_{req}}\right\}$, so the left $(1-f)$ of cooling demand must be met by mechanical chilling. Facility-level parasitic loads are summarized by a compact reduced-form term $W_{par}(L)$, capturing pumps and distribution losses that scale with system size and, potentially, distance. This representation preserves the key accounting identity for electric demand while avoiding engineering over-specification.

### B. PUE implications across IT load utilization

Figure 3 reports derived PUE as a function of IT load utilization. In the standalone benchmark, PUE exceeds unity because cooling power scales with IT load at rate $1/COP_m$, and fixed overheads create a disproportionate burden at low utilization. In the integrated system, delivered cooling is predominantly thermal; the residual electric burden is limited to parasitic loads and fixed overhead. Two implications follow.

First, the coupled configuration produces a lower PUE at all utilization levels considered, indicating that heat-driven cooling can reduce the marginal electricity intensity of computing even when generation efficiency in the WtE plant is not altered. Second, the PUE gap between standalone and coupled configurations is structurally bounded: once $f(L) = 1$(full thermal coverage), additional heat availability does not further reduce electric cooling demand, and PUE becomes governed by fixed overhead plus parasitics relative to IT load. This "coverage-induced plateau" is a recurring feature in subsequent sensitivity results.

### C. System-level quality-adjusted efficiency

Electricity-only efficiency is an incomplete welfare proxy for integrated energy–computing systems because the coupled configuration produces two valuable outputs: electricity and cooling (which offsets electricity that would otherwise be purchased for cooling). Figure 4 therefore evaluates systemic efficiency using a quality-adjusted accounting in which cooling is credited by an exergy factor that reflects its thermodynamic grade relative to ambient conditions. It shows that the integrated configuration attains higher systemic efficiency than the electricity-only WtE benchmark. The mechanism is grade matching: WtE's non-electric thermal output is low-grade relative to electricity but is well matched to the low-grade cooling demand of an AIDC campus. Converting heat to cooling therefore recovers economically valuable services from an energy stream that would otherwise be dissipated or under-utilized. Importantly, the improvement does not rely on optimistic assumptions about generation performance; it arises from reallocating the same primary energy input into a more appropriate output portfolio.

Figure 3. Performance sensitivity of PUE under variable IT load utilization: A comparative analysis of standalone and coupled architectures.

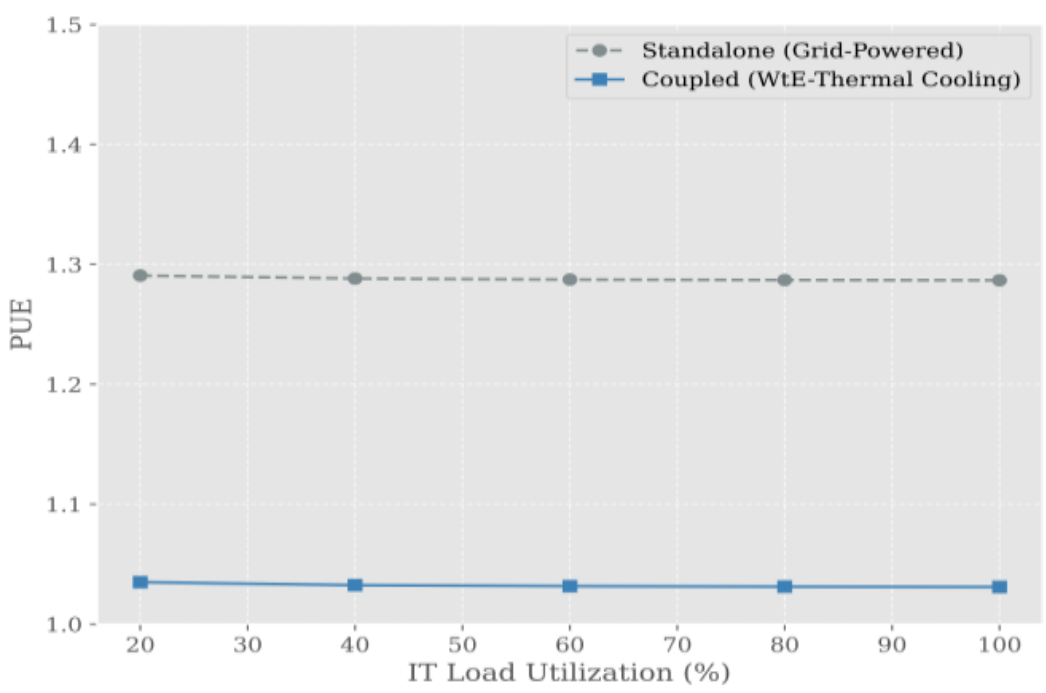

Figure 4. Comparative thermodynamic analysis of systemic exergy efficiency across standalone and integrated operational regimes.

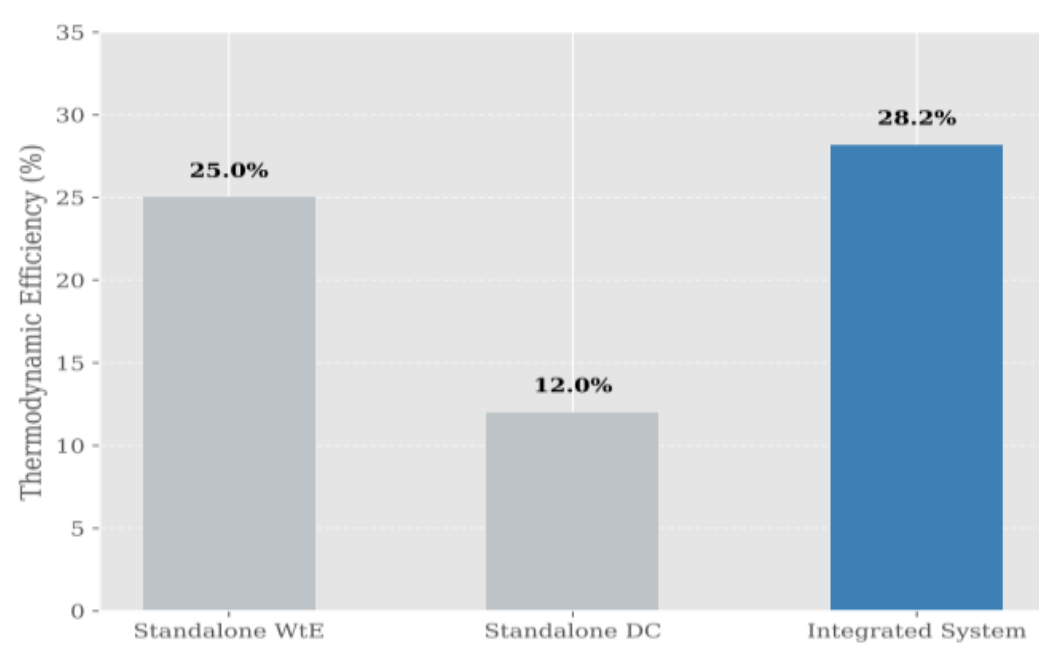

### D. *Feedstock sensitivity: $\boldsymbol{\Delta PUE_{elec}}$ vs waste LHV*

Figure 5 examines how the electric PUE gain, $\Delta PUE_{elec} \equiv PUE_{standalone} - PUE_{coupled}$, responds to feedstock lower heating value (LHV) under a fixed separation distance $L = 25$ km. Three utilization levels illustrate how load scale interacts with thermal supply.

Two patterns are robust. First, $\Delta PUE_{elec}$ is increasing in LHV in the partial-coverage region: higher-LHV waste increases available recoverable heat, raising $Q_{cool}(L)$ and thereby increasing $f(L)$. This reduces the electrically-driven share of cooling and lowers total facility electricity per unit IT. Second, $\Delta PUE_{elec}$ flattens once the system reaches full thermal coverage for the utilization scenario in question. The reference-case threshold $LHV^{\backslash *} \approx 7.91$ MJ/kg for $\rho = 60\%$ marks the point at which delivered absorption cooling meets the modeled cooling requirement; beyond this point, additional feedstock quality does not further reduce electric cooling demand, and gains are limited by parasitic and fixed overhead components.

The dependence on utilization is economically meaningful. Higher utilization requires a larger absolute cooling load, so full coverage is harder to achieve; as a result, the $\rho = 90\%$ curve exhibits a longer increasing segment before reaching its plateau. Conversely, the $\rho = 30\%$ scenario remains fully covered over the plotted LHV range under baseline assumptions, implying that marginal improvements in waste quality deliver limited incremental electric savings once coverage is already complete.

Figure 5. Cooling Coverage and Electricity Displacement under Feedstock Variability: Sensitivity to Waste LHV and Throughput

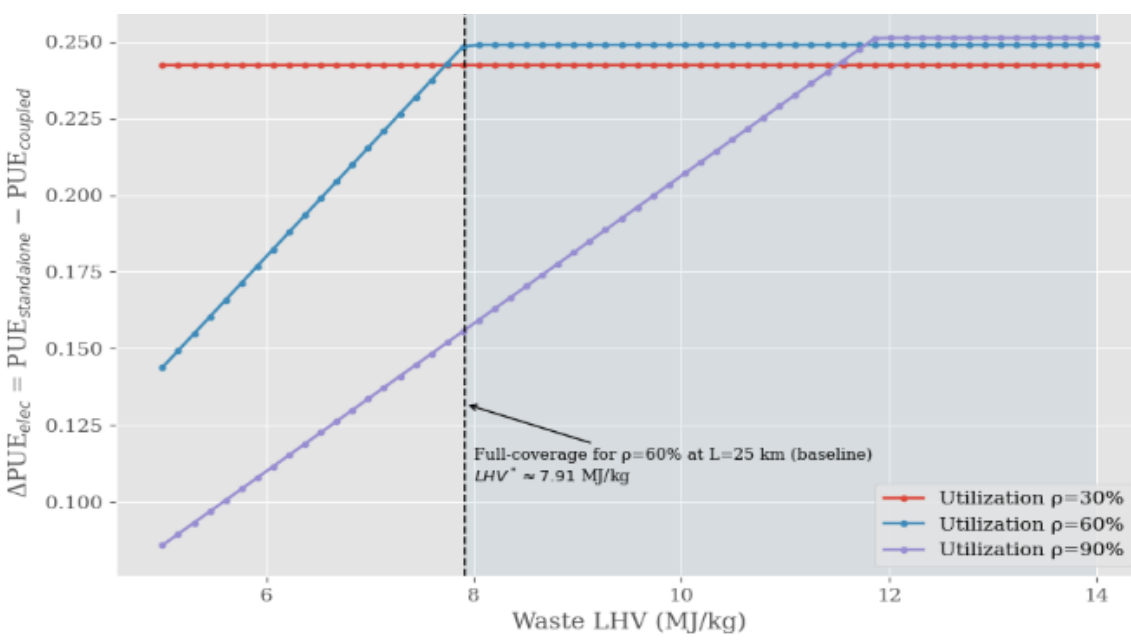

Figure 6. Net Benefit versus Corridor Distance: Thermal-Delivery Decay, Parasitic Work, and the Break-Even Radius for WtE–AIDC Coupling

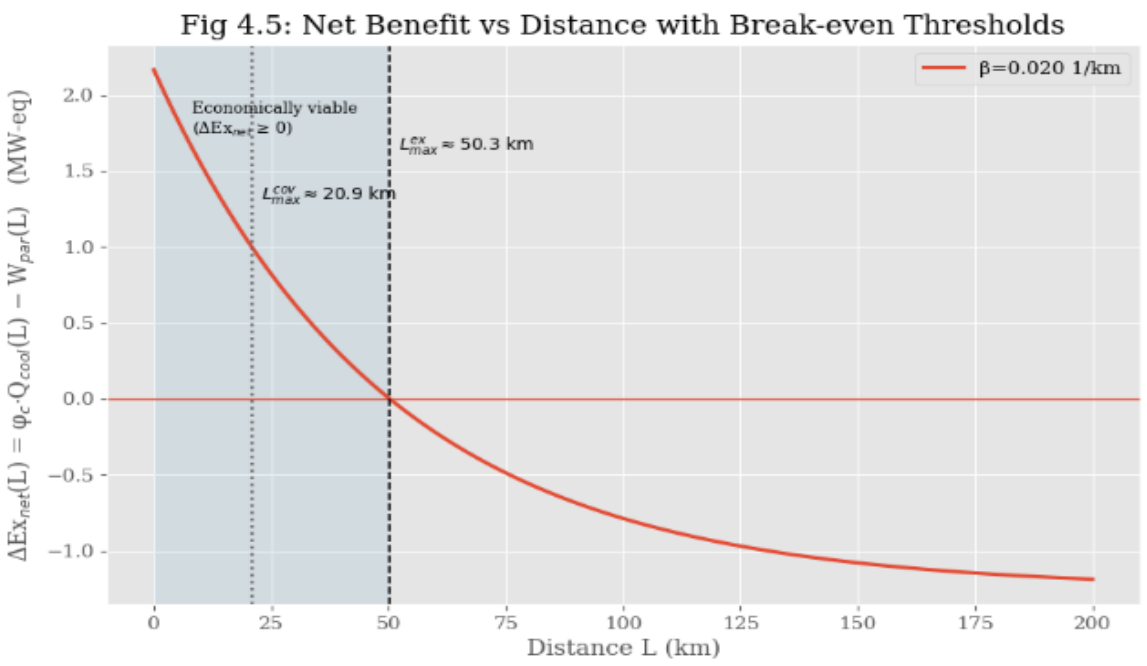

### E. *Spatial feasibility and break-even distances*

While feedstock quality governs thermal supply, spatial separation governs deliverability. Figure 6 characterizes distance constraints using a net exergy-benefit function, $\Delta Ex_{net}(L) \equiv \phi_c \, Q_{cool}(L) - W_{par}(L)$, where $\phi_c$ is the exergy factor associated with cooling at the chilled-water condition. The exponential decay in $\eta_{tr}(L)$ reduces delivered cooling with distance, while parasitics represent the electric cost of maintaining thermal transport and system operation.

Two thresholds emerge. The "coverage threshold" $L_{max}^{cov} \approx 20.9$km is the largest distance at which delivered absorption cooling can fully cover the modeled cooling requirement ($f(L) = 1$) in the reference configuration. Beyond $L_{max}^{cov}$, the system transitions to partial coverage, and mechanical chilling is required for the residual cooling demand. The "thermoeconomic break-even" $L_{max}^{ex} \approx 50.3$km is the largest distance at which the quality-adjusted cooling benefit remains non-negative after accounting for parasitic electricity. For $L \leq L_{max}^{ex}$, the coupled system is thermoeconomically viable in the sense that recovered cooling exergy weakly dominates its parasitic burden; for $L > L_{max}^{ex}$, attenuation and parasitics overwhelm the recovered value, and the coupling ceases to be justified on the modeled terms. These thresholds provide a compact siting implication: co-location or corridor-scale adjacency is not merely convenient but is often necessary for the coupling to deliver meaningful economic value. The results also clarify why distance is the central realism constraint in an otherwise favorable grade-matching environment, without short thermal links (or equivalently, without sufficiently low $\beta$ and parasitic intensity), recoverable heat cannot be monetized as reliable cooling services.

Across the four figures, three messages are consistent. First, coupling reduces electric intensity of computing and improves system-level quality-adjusted efficiency through grade matching. Second, the magnitude of PUE gains is governed by whether the system is in a partial-coverage or full-coverage regime; this produces observable plateaus in $\Delta PUE_{elec}$ once thermal coverage is achieved. Third, spatial decay creates hard feasibility boundaries, summarized by coverage and break-even distances, which translate the thermodynamic logic into actionable siting constraints. Together, these results motivate the subsequent discussion of levelized computing cost and strategic valuation under realistic corridor-scale deployments.

## V. Financial and Strategic Implications

### A. *Sustainability mechanisms and ESG boundary*

The ESG relevance of WtE–AIDC coupling follows from two measurable channels with clear counterfactual boundaries. First, diverting municipal solid waste from landfills to WtE alters the methane pathway. A disciplined reporting approach is to disclose diversion volume and composition $W_t$ and pair it with an explicit counterfactual (e.g., landfill with specified methane capture), then report emissions impacts as a sensitivity range using standardized factors rather than a single point estimate. Second, coupling reduces the electricity intensity of computing by displacing mechanical chilling. This can be reported directly via $\Delta PUE_{elec}$ (or normalized grid electricity $E_t^{grid}/K_t$) under stated utilization, feedstock, and distance assumptions. Importantly, Figures 5–6 imply these benefits are regime- and distance-dependent: claims should be conditioned on whether operation is within the viability window and in full versus partial cooling coverage.

### B. *Economic interface and decision metrics (LCOC, electricity displacement)*

To compare integrated designs on a consistent basis, we use a Levelized Cost of Computing (LCOC), $LCOC = \frac{\sum_t C_t(1+r)^{-t}}{\sum_t K_t(1+r)^{-t}}$, where $K_t$ can be selected based on data availability (conservatively $K_t = E_t^{IT}$, yielding \$/ $\text{MWh}_{IT}$). The coupling primarily affects purchased grid electricity through the coverage term $f(L)$:

$$E_t^{grid} = E_t^{IT} + E_t^{aux} + \frac{(1 - f(L))Q_{req,t}}{COP_m} + W_{par,t}(L) - E_t^{on\text{-}site} \quad (18)$$

### C. *Deployment (corridor strategy) and storage extension*

The comparative statics imply WtE–AIDC integration is best interpreted as a corridor strategy rather than a universal design rule: value is concentrated within limited co-location radii where deliverable heat can be monetized as reliable cooling at scale. Cold/thermal energy storage (CTES/TES) is a natural extension that buffers heat–cooling mismatch and can deepen electric-chiller displacement near the corridor frontier; the coverage and break-even conditions derived here provide first-order bounds for storage sizing and dispatch extensions.

## VI. Conclusions

This paper evaluates WtE-AIDC coupling as an energy-service substitution strategy: low-grade WtE heat drives absorption cooling to displace electricity-intensive mechanical chilling and reduce marginal grid demand. Using a transparent input–output boundary benchmarked against a standalone baseline, we show that thermoeconomic viability is governed by three first-order determinants, cooling coverage of IT heat load, parasitic auxiliary electricity, and distance-driven thermal delivery decay, which together define a siting-ready break-even corridor frontier. Sensitivities to utilization and feedstock quality follow the same regime logic, with benefits saturating once full thermal cooling coverage is reached. The framework provides decision-ready feasibility conditions and break-even thresholds for corridor-scale planning of sustainable AI infrastructure and can be extended to storage-enabled operation (TES/CTES) to buffer heat–cooling mismatch and broaden operating ranges.

Note: For the proof of Proposition 1 (Sufficient Condition for Thermoeconomic Superiority) in Section III, the derivations of comparative statics and threshold conditions in Section IV, the parameterization and scenario table, and the Python code used to produce the simulation design and results (Figs. 3–6), please refer to Online Appendix A–C and the accompanying GitHub repository.

## Acknowledgment

The views expressed in this paper are solely those of the author and do not necessarily reflect the views of Google LLC, and Dun & Bradstreet Inc, or any other current or former employer or affiliated institution.